\documentclass{article}
\usepackage[english]{babel}
\usepackage{braket}
\usepackage{mathtools}
\usepackage[letterpaper,top=2cm,bottom=2cm,left=3cm,right=3cm,marginparwidth=1.75cm]{geometry}

\usepackage{amsthm}

\usepackage{placeins}
\usepackage{xcolor}
\usepackage{listings}
\usepackage{biblatex} %Imports biblatex package

\usepackage{listings}
\usepackage{amsmath}
\usepackage{graphicx}
\usepackage[colorlinks=true, allcolors=blue]{hyperref}
\usepackage{algorithm}
\usepackage{algpseudocode}

\title{QAOA on Hamiltonian Cycle problem}
\author{Zhuoyang Ye, UCLA Physics and Astronomy, yezhuoyang98@g.ucla.edu}

\begin{document}
\maketitle

\begin{abstract}
I use QAOA to solve the Hamiltonian Circle problem. First, inspired by Lucas \cite{Lucas_2014}, I define the QUBO form of Hamiltonian Cycle an transform it to a quantum circuit by embedding the problem of $n$ vertices to an encoding of $(n-1)^2$ qubits. Then, I calcluate the spectrum of the cost hamiltonian for both triangle case and square case and justify my definition. I also write a python program to generate the cost hamiltonian automatically for finding the hamiltonian cycle in an arbitrary graph. I test the correctess of the hamailtonian \textbf{by analyze their energy spectrums}. Since the $(n-1)^2$ embedding limit my simulation of graph size to be less than $5$, I decide to test the correctness, only for small and simple graph in this project. I implement the QAOA algorithm using qiskit and run the simulation for the triangle case and the square case, which are easy to test the correctness, both with and without noise. A very interesting result I got is that \textbf{for the square case, the QAOA get much better result on a noisy simulator than a noiseless simulator}! The explanation for this phenomena require further investigation, perhaps quantum noise can actually be helpful, rather than harmful in the annealing algorithms. I also use two different kinds of mixer, $R_x$ mixer and $R_y$ circuit to run the simulation. It turns out that $R_x$ mixer performs much better than $R_y$ mixer in this problem.   
\end{abstract}

\section{Introduction}

QAOA, first introduce in 2014 \cite{farhi2014quantum}, is one of the the most famous and widely studied in NISQ era \cite{Preskill_2018} \cite{blekos2023review} of quantum computing. In 2020, Google AI quantum implement the QAOA on a real Sycamore superconducting qubit quantum processor \cite{Harrigan_2021}, where they, for the first time, got a non-trivial result using QAOA to solve maxcut problem, where the graph has the same topology as the real Hardware Grid. Recently, a group from Harvard used Rydberg atom arrays with up to 289 qubits in two spatial dimensions, and experimentally investigate quantum algorithms for solving the maximum independent set problem\cite{Lukin}. 

Despite all the effort and progress, whether QAOA has an advantage in solving classical intractable problem , especially NP-complete problem, is still an open question.

\section{Theory of QAOA}

QAOA is a typical variational quantum algorithm that could solve combinatorial optimization problem.

The initial idea of QAOA comes from quantum adabatic theorem. Consider a time dependent hamiltonian that evolve slowly with time. The initial hamiltonian is $\hat{H}_M$ and the final one after evolution time $T$ is $\hat{H}_C$. The adabitic theorem tells us that if we set the initial state as the ground state of $\hat{H}_M$, the final state will also be the ground state of $\hat{H}_C$. 

The theorem has a potential implementation, when we can easily prepare the ground state of $\hat{H}_C$, and encode the solution of a hard problem we want to solve to be the ground state of $\hat{H}_M$.  We can choose the $\hat{H}_M$, to be Pauli X gates on all qubits, the ground state of which, is simply $\ket{+}^{\otimes n}$, which can be easily prepared in a quantum computer by a row of Hadmard gate on the initial $\ket{0}^{\otimes n}$. 

The time dependent Hamiltonian can be expressed as
\begin{equation}
     \hat{H}(t)=f(t)\hat{H}_C + g(t)  \hat{H}_M
     \label{eq:AdabaticHamiltonian}
\end{equation}
The two function $f(t)$ and $g(t)$ changes slowly in time. An example for such $f(t)$ $g(t)$ is $f(t)=t/T$ and $g(t)=1-t/T$. The unitary of such slowly varying hamiltonian is 
\begin{equation}
     \hat{U}(t)= e^{-i \int_{0}^t d\tau \hat{H}(\tau)}
\end{equation}

 can be simulated in a quantum computer, by Trotterization:

\begin{equation}
    \hat{U}(t) \approx  \prod_{k=0}^{r-1} \exp[-i\hat{H}(k \Delta \tau)\Delta \tau]= \prod_{k=1}^{r-1} \exp[-if(k\Delta \tau)\hat{H}_C \Delta \tau ] \exp[-ig(k\Delta \tau)\hat{H}_M \Delta \tau]
    \label{eq:adiabatic}
\end{equation}

Next, let's talk about how to find suitable $\hat{H_C}$ with regard to the problem we want to solve and how to embed the solution to its ground state.

The most formal way to describe the classical problem that QAOA want to solve is define a set of boolean functions,  $\{C_{\alpha}, \alpha=0,1,\cdots,m\}$, each one of these function define a condition to be satisfied given the specific problem, or mathematically as a mapping: $C_\alpha: \{ 0,1\}^n \rightarrow \{0,1\}$. The input of the boolean function represent an ``assignment'' to the given problem and the output is whether the assignment belong to the solution space or not. 

The optimization version of the problem, when there are ,can be stated in the function below, formally as:
\begin{equation}
    C(z)=\sum_{\alpha=1}^m C_{\alpha} (z)
    \label{eq:Cost}
\end{equation}

When $C(z)=m$, all of the conditions of the problem are satisfied, otherwise some of them are not. Our goal is to maximized the value of $C(z)$. 

The way to construct the circuit of QAOA, is to make alternating layers of mixer and cost circuit that could possibly simulate the adiabatic evolution of quantum hamiltonian in euqation \hyperref[eq:adiabatic]{Equation (3)}. 

However, if we want to get accurate solution, we have to choose infinitly small $\Delta \tau$ in \hyperref[eq:adiabatic]{Equation (3)}, which is unacceptable for a  quantum computer in the real application. Thus, in QAOA algorithm, we only construct a fixed number of layers, and assign each layer with a parameter to be optimized, under the hope that after such optimization, the hamiltonian of the quantum computer approximate the Trotterized adiabitic unitary very well. 

The optimization algorithm itself, utilize the similar idea from training a neural network. The parameters are changed every time we execute the circuit and measured the result. People also call this Variational Quantum Eigenvalue (VQE) algorithms \cite{VQE}. The same kind of methods have demonstrated it's potential in solving eigenstate and energy for chemical molecule. A very recent result is Google run VQE on 12 qubits on Sycamore \cite{GoogleVQE} and get the ground state energy for hydrogen chains.

\subsection{Structure of Cost circuit}

First, we have to design the cost hamiltonian $\hat{H}_C$ with regard to the given problem.  All of the possible solutions for embedded as $\ket{x}$ should be an eigen state of $\hat{H}_C$, whose eigen value contain the information of the cost function: 
\begin{equation}
      \hat{H}_C \ket{x} =C(x) \ket{x}
      \label{eq:CostHamiltonian}
\end{equation}

The unitary for the cost circuit, with parameter $\gamma$ , is defined as
\begin{equation}
    U_C(\gamma)=e^{-i\gamma \hat{H}_C} = \prod_{j<k} e^{-i\gamma w_{jk} \hat{Z}_j\hat{Z}_k}
    \label{eq:unicost}
\end{equation}

The circuit, can be implemented by two CNOT gate and a $R(Z)$ gate

\subsection{Structure of Mixer circuit}

The mixer hamiltonion, is choosen to be 
\begin{equation}
      \hat{H}_M=\sum_{j \in \mathcal{V}} \hat{X}_j
      \label{eq:Mixer}
\end{equation}

The unitary for the mixer part, with parameter $\beta$, is defined as 
\begin{equation}
    U_M(\beta)=e^{-i\beta B}= \prod_j e^{-i \beta \hat{X}_j}
    \label{eq:unmixer}
\end{equation}
The circuit, can be implemented by a $R_X$ gate.

We can also choose another mixer hamiltonion, such as Grover Mixer \cite{GroverMixer},

\subsection{Structure of the QAOA circuit}
Just as most of the circuti structure of quantum algorithm, there should be a row of Hadamard gate ad the front of the circuit that convert $\ket{0}^{\otimes n}$ to  $\ket{+}^n$. And then, we add $p$ alternating layers of Cost circuit and Mixer circuit. The two set of parameters, are denoted as $\gamma=(\gamma_1,\cdots,\gamma_p)$ and  $\beta=(\beta_1,\cdots,\beta_p)$. The  

\begin{equation}
    \ket{\gamma,\beta}=U_M(\beta_p) U_C(\gamma_p) \dots U_M(\beta_1) U_C(\gamma_1) \ket{+}^{\otimes n}
    \label{eq:Equation}
\end{equation}

And the measured cost function of the equation \hyperref[eq:Equation]{equation (4)}. 

\begin{equation}
     \braket{C}= \bra{\gamma, \beta} \hat{H}_C \ket{\gamma, \beta}
\end{equation}

\subsection{Optimization algorithm}

We can simply use a classical optimizer to optimize all the parameters.

\subsection{Steps of the algorithm}

\begin{enumerate}
    \item Define a cost Hamiltonion $\hat{H}_C$ given the problem. The eigen state with the highest eigen energy of $\hat{H}_C$ should be the exact solution to the optimization problem.
    \item Initialize the state in $\ket{s}$. 
    \begin{equation}
          \ket{s}= \ket{+}^{\otimes n}=\frac{1}{\sqrt{2^n}} \sum_{x \in \{0,1\}^n} \ket{x}
          \label{eq:S}
    \end{equation} $\ket{s}$ here is actually the eigen state of the highest eigen state of the mixer hamiltonian $H_M$ defined in \hyperref[eq:Mixer]{Equation (7)} 
    \item Choose the number of layer $p$. Make $p$ alternating pair of mixer and cost circuit. 
    \item Initialize $2p$ parameters $\vec{\gamma}=(\gamma_1,\gamma_2,\cdots,\gamma_p)$ and $\vec{\beta}=(\beta_1,\beta_2,\cdots,\beta_p)$ such that $\gamma_i,\beta_k \in \{0,2\pi \}$
    \item Calculate the cost by measuring repeatedly. 
    \begin{equation}
           F_p(\vec{\gamma},\vec{\beta})=\bra{\psi_p(\vec{\gamma},\vec{\beta})}H_C \ket{\psi_p(\vec{\gamma},\vec{\beta})} 
           \label{eq:Fidelity}
    \end{equation}
    \item Use a classical algorithm to optimize the paramter by maximize the expectation value in \hyperref[eq:Fidelity]{Equation (12)}.
    \begin{equation}
               (\vec{\gamma}^*,\vec{\beta}^*)= \arg \max_{\vec{\gamma},\vec{\beta}}  F_p(\vec{\gamma},\vec{\beta})
               \label{eq:ClassicalOp}
    \end{equation}
 
\end{enumerate}

\subsection{MAX-CUT}

I use MAX cut problem first to test the  QAOA circuit. Which is the most commonly used algorithm to benchmark the behavior of QAOA. 

The input to the MAXCUT is a graph $\mathcal{G}=(\mathcal{V},\mathcal{E})$. $\mathcal{V}$ is the set of vertices, $\mathcal{E}$ is the set of edges. Assume that there is a weight $w_{ij}$ assigned to each of the edge $(i,j) \in \mathcal{E}$. We want to find the largest cut in graph $\mathcal{G}$, which is a subset of the vertices whose ``cut'' with the rest of the vertices are the maximum. The ``cut'' is defined as the sum of all the weight between a vertex in the subset and a vertex that is not in the subset. Any possible assignment can be represented by a set of $0,1$. So we use $x_i$ to represent the assignment for vertex $i$. $x_i=1$ if and only if $x_i$ is assigned to the subset.

\begin{equation}
     C(x)=\sum_{i,j=1}^{|\mathcal{V}|} w_{i,j}x_i(1-x_j)
     \label{eq:Cmax}
\end{equation}

The correspondence between the $x_i$ in \hyperref[eq:Cmax]{Equation (11)} with the Pauli Z gate used in our Cost Hamiltonian is:
\begin{equation}
     x_i \rightarrow \frac{1}{2} (1- Z_i)
     \label{eq:xToZ}
\end{equation}

For example, for the cost function of MAX-CUT defined in  \hyperref[eq:Cmax]{Equation (11)}, the corresponding cost hamiltonian is 
\begin{equation}
    H_C= \sum_{i,j=1}^{|\mathcal{V}|} w_{i,j}    \frac{1}{4}   (1-Z_i)Z_j= \frac{1}{4}   \sum_{i,j=1}^{|\mathcal{V}|} w_{i,j} (Z_j-Z_iZ_j)
    \
\end{equation}

\section{Classical NP-complete problem}

In 1971, \cite{Cook} Cook first proved that the boolean satisfiability problem(3-SAT) is NP-complete, which is also called the Cook-Levin theorem. In 1972, Karp \cite{Karp} used Cook's result and first introduce 21 famous NP complete problem.  In 2014, Lucas \cite{Lucas_2014} first discussed how to map all of the 21 NP complete to Quadratic Unconstrained Binary Optimization problems(QUBO) in polynomial time, which suddenly raise people's attention because this open the door for using the quantum computer to solve NP-complete problem.

One of the most notorious NP complete problem is the Hamiltonian Circle problem. I will try to solve the problem using QAOA in this small project.

\subsection{Hamiltonian Circle problem}

The input for the Hamiltonian Circle problem is a graph $\mathcal{G}=(\mathcal{V},\mathcal{E})$. Suppose $|\mathcal{G}|=n$. Our goal is to find a cycle to travel thorough all vertices exactly once. I use the QUBO form given by Lucas \cite{Lucas_2014} for the Hamiltonian Circle as follows:

\begin{equation}
     H=A \sum_{v=1}^n(1-\sum_{j=1}^nx_{v,j})^2+A \sum_{j=1}^n(1-\sum_{v=1}^nx_{v,j})^2+A \sum_{(uv) \notin \mathcal{E}} [\sum_{j=1}^{n-1} x_{u,j}x_{v,j+1} + x_{u,n}x_{v,1}]
     \label{eq:HCircle}
\end{equation}

$x_{v,j}$ is the encoding of whether the vertex $v$ is at the $j^{th}$ position of the Circle. The first part of $H$ requires that every vertex can be only assigned to one position in the Circle, which we call \textbf{vertex uniqueness term}. The second part is the constraint that every position in the Circle of length $N$ we find is only assigned to one vertex, which we call \textbf{edge uniqueness term}. The final term is the penalty for the two consecutive vertices in the Circle are actually not connected in the original graph, which we call \textbf{edge validity term}. Thus, it is obvious that the minimal value of $H$, given any assignment $x_{v,j}=f(v,j)$, is $0$, when such assignment represent a hamiltonian Circle. 

To construct the circuit, we can also use the following substitution:
\begin{equation}
     x_{v,j} \rightarrow \frac{1}{2} (1- Z_{i,j})
\end{equation}

The encoding of a given graph requires $(n-1) \times (n-1)$ qubits.

\begin{figure}[h!]
    \centering
    \includegraphics[width=0.4\linewidth]{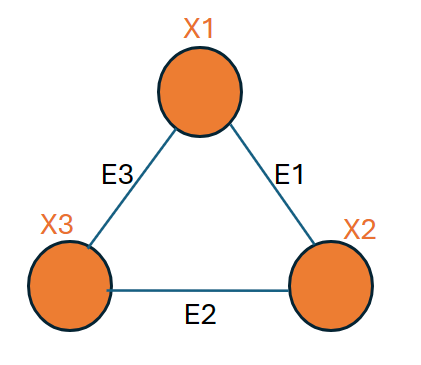}
    \caption{The sketch of a triangle which we want to find a hamiltonian circle.}
    \label{fig:circle3}
\end{figure}

First, we calculate a very simple case, a triangle, as illustrate in \autoref{fig:circle3}.  

In this simple example, we can write the full embedding in details:

\begin{equation}
\begin{aligned}
      &x_{1,1}\equiv 1 \quad \text{We always fix the position of the first vertex to be 1.} \\
    &x_{1,2}\equiv x_{1,3}\equiv x_{2,1}\equiv x_{3,1}\equiv 0 \quad \text{The impossible assignment when the first vertex is fixed.} \\
      &x_{2,2}=1 \quad \text{Vertex 2 is at the second position.} \\
      &x_{2,3}=1 \quad \text{Vertex 2 is at the third position.} \\      
      &x_{3,3}=1 \quad \text{Vertex 2 is at the third position.} \\    
\end{aligned}    
\end{equation}

By the above definition, we can see that $2 \times 2$ qubits are enough for defining the hamiltonian for hamiltonian circle.

Now we can derive the concrete form of the hamiltonian defined in \autoref{eq:HCircle}, where we set the constant $A=1$.

\begin{equation}
     H=\sum_{v=2}^3(1-\sum_{j=2}^3x_{v,j})^2+ \sum_{j=2}^3(1-\sum_{v=2}^3x_{v,j})^2+[\sum_{(uv) \notin \mathcal{E}} \sum_{j=2}^2 x_{u,j}x_{v,j+1} + \sum_{(u1) \notin \mathcal{E}}x_{u,3}+\sum_{(1u) \notin \mathcal{E}}x_{u,2}]
     \label{eq:HCircle}
\end{equation}
We expand the three terms seperately

\par 1.\textbf{The vertex uniqueness term}:

\begin{align*}
  H_1 &=\sum_{v=2}^3(1-\sum_{j=2}^3x_{v,j})^2\\
      &=(1-x_{2,2}-x_{2,3})^2+(1-x_{3,2}-x_{3,3})^2\\
      &=2+x_{2,2}^2+x_{2,3}^2+x_{3,2}^2+x_{3,3}^2-2x_{2,2}-2x_{2,3}+2x_{2,2}x_{2,3}-2x_{3,2}-2x_{3,3}+2x_{3,2}x_{3,3}\\
\end{align*}

We can simply use the substitution rule $x_{i,j} \rightarrow \frac{1}{2}(1-Z_{i,j})$

\begin{align*}
  H_1 &=\sum_{v=2}^3(1-\sum_{j=2}^3x_{v,j})^2\\
      & \Rightarrow (\frac{1}{2}Z_{2,2}+\frac{1}{2}Z_{2,3})^2+(\frac{1}{2}Z_{3,2}+\frac{1}{2}Z_{3,3})^2\\
      &=\frac{1}{4}[I+I+2Z_{2,2}Z_{2,3}+I+I+2Z_{3,2}Z_{3,3}]\\
      &=I+\frac{1}{2}Z_{2,2}Z_{2,3}+\frac{1}{2}Z_{3,2}Z_{3,3}
\end{align*}

\par 2.\textbf{The edge uniqueness term}:
\begin{align*}
H_2&=\sum_{j=2}^3(1-\sum_{v=2}^3x_{v,j})^2\\
   &=(1-x_{2,2}-x_{3,2})^2+(1-x_{2,3}-x_{3,3})^2\\
\end{align*}

We also use the substitution rule $x_{i,j} \rightarrow \frac{1}{2}(1-Z_{i,j})$

\begin{align*}
  H_2 &=(1-x_{2,2}-x_{3,2})^2+(1-x_{2,3}-x_{3,3})^2\\
      & \Rightarrow (\frac{1}{2}Z_{2,2}+\frac{1}{2}Z_{3,2})^2+(\frac{1}{2}Z_{2,3}+\frac{1}{2}Z_{3,3})^2 \\
      & =\frac{1}{4}(I+I+2Z_{2,2}Z_{3,2}+I+I+2Z_{2,3}Z_{3,3})\\
      & = I +\frac{1}{2}Z_{2,2}Z_{3,2}+\frac{1}{2}Z_{2,3}Z_{3,3}
\end{align*}

\par 3.\textbf{The edge validity term}:
\[
H_3=\sum_{(uv) \notin \mathcal{E}} \sum_{j=2}^2 x_{u,j}x_{v,j+1} + \sum_{(u1) \notin \mathcal{E}}x_{u,3}+\sum_{(1u) \notin \mathcal{E}}x_{u,2}
\]
Notice the the triangle is actually a completely connected graph, so $H_3=0$.
Finally, we have

\begin{align*}
         H_C &=H_1+H_2+H_3\\
           &=2I+\frac{1}{2}Z_{2,2}Z_{2,3}+\frac{1}{2}Z_{3,2}Z_{3,3}+\frac{1}{2}Z_{2,2}Z_{3,2}+\frac{1}{2}Z_{2,3}Z_{3,3}
\end{align*}

Since I and constant factor don't affect the eigen energy and eigen state, we can rewrite $H_C$ as

\begin{align*}
         H_C &=Z_{2,2}Z_{2,3}+Z_{3,2}Z_{3,3}+Z_{2,2}Z_{3,2}+Z_{2,3}Z_{3,3}
\end{align*}

Finally, we can construct the circuit for the cost hamiltonian. The circuit has four qubits, each represent:
\begin{center}
  \begin{enumerate}
     \item $Q_1$ represent $x_{2,2}$.
     \item $Q_2$ represent $x_{2,3}$.
     \item $Q_3$ represent $x_{3,2}$.
     \item $Q_4$ represent $x_{3,3}$.
\end{enumerate}  
\end{center}
Since the correct result must be either $(x_{2,2}=1,x_{2,3}=0,x_{3,2}=0,x_{3,3}=1)$, which represent the hamiltonian cycle $(1 \rightarrow 2\rightarrow3\rightarrow1 )$ or $(x_{2,2}=0,x_{2,3}=1,x_{3,2}=1,x_{3,3}=0)$, which represent the hamiltonian cycle $(1 \rightarrow 3\rightarrow2\rightarrow1 )$. The ground state for this solution must in dirac notation be either $\ket{1001}$ or $\ket{0110}$. We can check the correctness.
We can write our $H_C$ as: 
\begin{equation}
  \begin{aligned}
         H_C &=Z_1Z_2+Z_3Z_4+Z_1Z_3+Z_2Z_4
\end{aligned}
\label{eq:HamiltonianTriangle}
\end{equation}

Let's check the correctness of the above statement by diagonalization:
\begin{lstlisting}
import numpy as np
from scipy.linalg import eigh
from functools import reduce
import matplotlib.pyplot as plt

# Pauli Z matrix
pauli_z = np.array([[1, 0], [0, -1]])

# Function to create a matrix representation of Z_k gate on k-th qubit
def z_k_matrix(k, total_qubits):
    I = np.eye(2)  # Identity matrix
    matrices = [pauli_z if i == k else I for i in range(total_qubits)]
    return reduce(np.kron, matrices)

# Terms for the Hamiltonian H_C
terms_direct = [
    (1, [0, 1]),  # Z1Z2
    (1, [2, 3]),  # Z3Z4
    (1, [0, 2]),  # Z1Z3
    (1, [1, 3])   # Z2Z4
]

# Total number of qubits for the Hamiltonian
new_total_qubits = 4

# Construct the Hamiltonian matrix directly
H_direct = np.zeros((2**new_total_qubits, 2**new_total_qubits))

# Add each term directly to the Hamiltonian
for coeff, qubits in terms_direct:
    term = np.eye(2**new_total_qubits)
    for qubit in qubits:
        term = np.dot(term, z_k_matrix(qubit, new_total_qubits))
    H_direct += coeff * term

# Calculate eigenvalues and eigenvectors of the Hamiltonian directly
new_eigenvalues_direct, new_eigenvectors_direct = eigh(H_direct)

# Plot the energy spectrum with annotations
plt.figure(figsize=(12, 8))
previous_eigenvalue = None
offset_multiplier = 0

for i in range(2**new_total_qubits):
    eigenvalue = new_eigenvalues_direct[i]
    max_amplitude_index = np.argmax(np.abs(new_eigenvectors_direct[:, i]))
    dirac_state = "|{0:04b}>".format(max_amplitude_index)
    if previous_eigenvalue == eigenvalue:
        offset_multiplier += 1
    else:
        offset_multiplier = 0
    horizontal_position =  + offset_multiplier * 0.088
    plt.hlines(eigenvalue, 0, 1, colors='b', linestyles='solid')
    plt.text(horizontal_position, eigenvalue, dirac_state, fontsize=12, 
        verticalalignment='center')
    previous_eigenvalue = eigenvalue

plt.xlabel('State Index',fontsize=20)
plt.ylabel('Energy',fontsize=20)
plt.title('Energy Spectrum and Corresponding Eigenstates 
    (Dirac Notation)',fontsize=20)
plt.xticks([])
plt.grid(True)
plt.savefig("SpectrumTriangle.png")
plt.show()
\end{lstlisting}

\begin{figure}[h!]
    \centering
    \includegraphics[width=\linewidth]{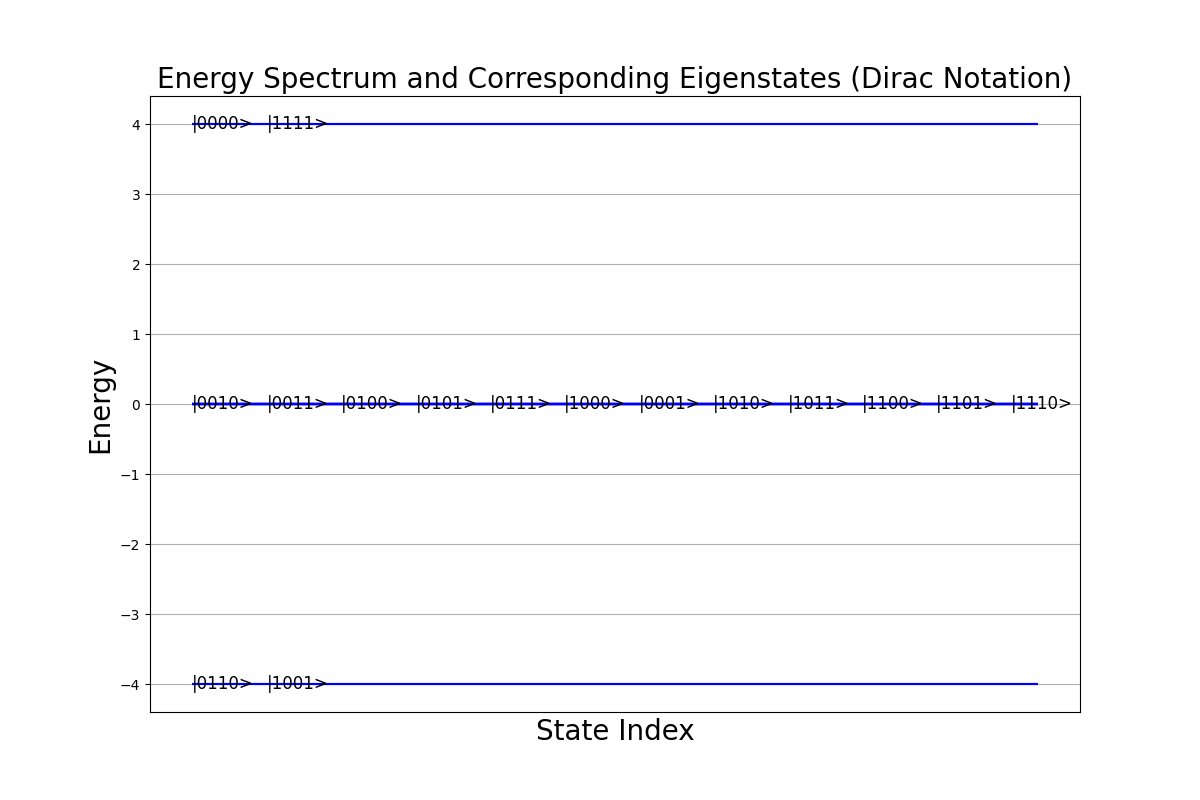}
    \caption{The sketch of the cost function for the triangle. The lowest energy is $-4$, with eigenstates $\ket{0110}$ and $\ket{1001}$, which are the exact solution for hamiltonian cycle for a triangle. The spectrum justify the correctness of our definition.}
    \label{fig:TriangleSpec}
\end{figure}

The spectrum of the hamiltonian calculated above is computed and shown in \autoref{fig:TriangleSpec}. There is a large energy gap, between the solution state and the non-solution state.

\FloatBarrier

Another example we is a square , as illustrate in \autoref{fig:circle4}:  

\begin{figure}[h!]
    \centering
    \includegraphics[width=0.4\linewidth]{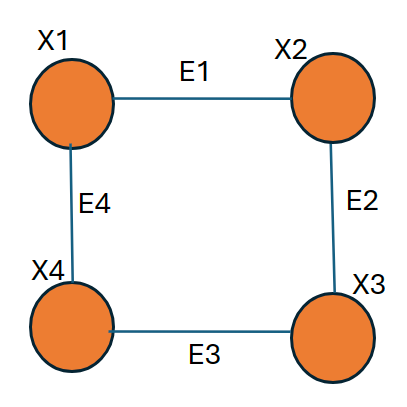}
    \caption{The sketch of a rectangle from where we want to find a hamiltonian circle.}
    \label{fig:circle4}
\end{figure}

In this simple example, we can also write the full embedding in details:

\begin{equation}
\begin{aligned}
      &x_{1,1}\equiv 1 \quad \text{We always fix the position of the first vertex to be 1.} \\
    &x_{1,2}\equiv x_{1,3}\equiv x_{1,4} \equiv x_{2,1} \equiv x_{3,1} \equiv x_{4,1} \equiv 0 \quad \text{The impossible assignment when the first vertex is fixed.} \\
      &x_{2,2}=1 \quad \text{Vertex 2 is at the second position.} \\
      &x_{2,3}=1 \quad \text{Vertex 2 is at the third position.} \\
      &x_{2,4}=1 \quad \text{Vertex 2 is at the fourth position.} \\      
      &x_{3,2}=1 \quad \text{Vertex 3 is at the second position.} \\ 
      &x_{3,3}=1 \quad \text{Vertex 2 is at the third position.} \\    
      &x_{3,4}=1 \quad \text{Vertex 2 is at the fourth position.} \\    
      &x_{4,2}=1 \quad \text{Vertex 2 is at the third position.} \\  
      &x_{4,3}=1 \quad \text{Vertex 2 is at the third position.} \\   
      &x_{4,4}=1 \quad \text{Vertex 2 is at the third position.} \\   
\end{aligned}    
\end{equation}

We can see that $3 \otimes 3=9$  for defining the hamiltonian for hamiltonian circle.

Now we can derive the concrete form of the hamiltonian defined in \autoref{eq:HCircle}, where we set the constant $A=1$.

\begin{equation}
     H=\sum_{v=2}^4(1-\sum_{j=2}^4x_{v,j})^2+ \sum_{j=2}^4(1-\sum_{v=2}^4x_{v,j})^2+\sum_{(uv) \notin \mathcal{E}} \sum_{j=2}^4 x_{u,j}x_{v,j+1}
     \label{eq:HCircle2}
\end{equation}
We expand the four terms seperately

\par 1.\textbf{The vertex uniqueness term}:

\begin{align*}
  H_1 &=\sum_{v=2}^4(1-\sum_{j=2}^4x_{v,j})^2\\
      &=(1-x_{2,2}-x_{2,3}-x_{2,4})^2+(1-x_{3,2}-x_{3,3}-x_{3,4})^2+(1-x_{4,2}-x_{4,3}-x_{4,4})^2\\
\end{align*}

We can simply use the substitution rule $x_{i,j} \rightarrow \frac{1}{2}(1-Z_{i,j})$

\begin{align*}
  H_1 &=\sum_{v=2}^4(1-\sum_{j=2}^4x_{v,j})^2\\
      & =(-\frac{I}{2}+\frac{1}{2}Z_{2,2}+\frac{1}{2}Z_{2,3}+\frac{1}{2}Z_{2,4})^2+(-\frac{I}{2}+\frac{1}{2}Z_{3,2}+\frac{1}{2}Z_{3,3}+\frac{1}{2}Z_{3,4})^2+(-\frac{I}{2}+\frac{1}{2}Z_{4,2}+\frac{1}{2}Z_{4,3}+\frac{1}{2}Z_{4,4})^2\\
      &=\frac{1}{4}[I+Z_{2,2}^2+Z_{2,3}^2+Z_{2,4}^2-2Z_{2,2}-2Z_{2,3}-2Z_{2,4}+2Z_{2,2}Z_{2,3}+2Z_{2,2}Z_{2,4}+2Z_{2,3}Z_{2,4}+  \\
            &I+Z_{3,2}^2+Z_{3,3}^2+Z_{3,4}^2-2Z_{3,2}-2Z_{3,3}-2Z_{3,4}+2Z_{3,2}Z_{3,3}+2Z_{3,2}Z_{3,4}+2Z_{3,3}Z_{3,4}+ \\
            &I+Z_{4,2}^2+Z_{4,3}^2+Z_{4,4}^2-2Z_{4,2}-2Z_{4,3}-2Z_{4,4}+2Z_{4,2}Z_{4,3}+2Z_{4,2}Z_{4,4}+2Z_{4,3}Z_{4,4}]\\
      &=\frac{1}{4}[12I-2Z_{2,2}-2Z_{2,3}-2Z_{2,4}+2Z_{2,2}Z_{2,3}+2Z_{2,2}Z_{2,4}+2Z_{2,3}Z_{2,4} \\
                   &-2Z_{3,2}-2Z_{3,3}-2Z_{3,4}+2Z_{3,2}Z_{3,3}+2Z_{3,2}Z_{3,4}+2Z_{3,3}Z_{3,4}+   \\
                   &-2Z_{4,2}-2Z_{4,3}-2Z_{4,4}+2Z_{4,2}Z_{4,3}+2Z_{4,2}Z_{4,4}+2Z_{4,3}Z_{4,4}]   
\end{align*}

Finally, after removing $I$ and the constant factor:

\begin{equation}
\begin{aligned}
  H_1 &=-Z_{2,2}-Z_{2,3}-Z_{2,4}+Z_{2,2}Z_{2,3}+Z_{2,2}Z_{2,4}+Z_{2,3}Z_{2,4} \\
                   &-Z_{3,2}-Z_{3,3}-Z_{3,4}+Z_{3,2}Z_{3,3}+Z_{3,2}Z_{3,4}+Z_{3,3}Z_{3,4}+   \\
                   &-Z_{4,2}-Z_{4,3}-Z_{4,4}+Z_{4,2}Z_{4,3}+Z_{4,2}Z_{4,4}+Z_{4,3}Z_{4,4}       
\end{aligned}
\end{equation}

\par 2.\textbf{The edge uniqueness term}:
\begin{align*}
H_2&=\sum_{j=2}^4(1-\sum_{v=2}^4x_{v,j})^2\\
   &=(1-x_{3,2}-x_{3,2}-x_{4,2})^2+(1-x_{2,2}-x_{2,3})^2\\
\end{align*}

We also use the substitution rule $x_{i,j} \rightarrow \frac{1}{2}(1-Z_{i,j})$

\begin{align*}
  H_2 &=(1-x_{2,2}-x_{3,2}-x_{4,2})^2+(1-x_{2,3}-x_{3,3}-x_{4,3})^2+(1-x_{2,4}-x_{3,4}-x_{4,4})^2\\
\end{align*}

Likewise, we can use the substitution rules:

We also also use the substitution rule $x_{i,j} \rightarrow \frac{1}{2}(1-Z_{i,j})$, and finally get the same kind of format as $H_1$:

\begin{equation}
\begin{aligned}
  H_2 &=-Z_{2,2}-Z_{3,2}-Z_{4,2}+Z_{2,2}Z_{3,2}+Z_{2,2}Z_{4,2}+Z_{3,2}Z_{4,2} \\
                   &-Z_{2,3}-Z_{3,3}-Z_{4,3}+Z_{2,3}Z_{3,3}+Z_{2,3}Z_{4,3}+Z_{3,3}Z_{4,3}+   \\
                   &-Z_{2,4}-Z_{3,4}-Z_{4,4}+Z_{2,4}Z_{3,4}+Z_{2,4}Z_{4,4}+Z_{3,4}Z_{4,4}       
\end{aligned}
\end{equation}

\par 3.\textbf{The edge validity term}:

\[ 
    H_3=\sum_{(uv) \notin \mathcal{E}} \sum_{j=2}^3 x_{u,j}x_{v,j+1} + \sum_{(u1) \notin \mathcal{E}}x_{u,4}+\sum_{(1u) \notin \mathcal{E}}x_{u,2}
\]

Different from the triangle case, there are two pairs of vertices not connected with each other: $(X_1,X_3)$ and $(X_2,X_4)$. Which means that our cost function will punish the assignment which try to find a path with $X_1, X_3$ or $X_2,X_4$ adjacent with each other.

\[
    H_3=x_{2,2}x_{4,3}+x_{2,3}x_{4,4}+x_{4,2}x_{2,3}+x_{4,3}x_{2,4}+x_{3,4}+x_{3,2}
\]

After substitution, the edge validity term becomes:

\begin{align*}
    H_3&=(I-Z_{2,2})(I-Z_{4,3})+(I-Z_{2,3})(I-Z_{4,4})+(I-Z_{4,2})(I-Z_{2,3})\\
    &+(I-Z_{4,3})(I-Z_{2,4})+2(I-Z_{3,4})+2(I-Z_{3,2})   \\
    &=-Z_{2,2}-Z_{4,3}+Z_{2,2}Z_{4,3}-Z_{2,3}-Z_{4,4}+Z_{2,3}Z_{4,4}-Z_{4,2}-\\
    &Z_{2,3}+Z_{4,2}Z_{2,3}-Z_{4,3}-Z_{2,4}+Z_{4,3}Z_{2,4}-2Z_{3,4}-2Z_{3,2}\\
    &=-Z_{2,2}-2Z_{4,3}-2Z_{2,3}-Z_{4,4}-Z_{4,2}-Z_{2,4}-2Z_{3,4}-2Z_{3,2}\\
    &+Z_{2,2}Z_{4,3}+Z_{2,3}Z_{4,4}+Z_{4,2}Z_{2,3}+Z_{4,3}Z_{2,4}
\end{align*}

\begin{align*}
         H_C &=H_1+H_2+H_3\\
           &=-Z_{2,2}-Z_{2,3}-Z_{2,4}+Z_{2,2}Z_{2,3}+Z_{2,2}Z_{2,4}+Z_{2,3}Z_{2,4} \\
                   &-Z_{3,2}-Z_{3,3}-Z_{3,4}+Z_{3,2}Z_{3,3}+Z_{3,2}Z_{3,4}+Z_{3,3}Z_{3,4}+   \\
                   &-Z_{4,2}-Z_{4,3}-Z_{4,4}+Z_{4,2}Z_{4,3}+Z_{4,2}Z_{4,4}+Z_{4,3}Z_{4,4}  
\end{align*}

\begin{equation}
\begin{aligned}
  H_C &=-Z_{2,2}-Z_{2,3}-Z_{2,4}+Z_{2,2}Z_{2,3}+Z_{2,2}Z_{2,4}+Z_{2,3}Z_{2,4} \\
                   &-Z_{3,2}-Z_{3,3}-Z_{3,4}+Z_{3,2}Z_{3,3}+Z_{3,2}Z_{3,4}+Z_{3,3}Z_{3,4}+   \\
                   &-Z_{4,2}-Z_{4,3}-Z_{4,4}+Z_{4,2}Z_{4,3}+Z_{4,2}Z_{4,4}+Z_{4,3}Z_{4,4}\\
                 &  -Z_{2,2}-Z_{3,2}-Z_{4,2}+Z_{2,2}Z_{3,2}+Z_{2,2}Z_{4,2}+Z_{3,2}Z_{4,2}\\
                   &-Z_{2,3}-Z_{3,3}-Z_{4,3}+Z_{2,3}Z_{3,3}+Z_{2,3}Z_{4,3}+Z_{3,3}Z_{4,3}+   \\
                   &-Z_{2,4}-Z_{3,4}-Z_{4,4}+Z_{2,4}Z_{3,4}+Z_{2,4}Z_{4,4}+Z_{3,4}Z_{4,4}\\
                   &-Z_{2,2}-2Z_{4,3}-2Z_{2,3}-Z_{4,4}-Z_{4,2}-Z_{2,4}-2Z_{3,4}-2Z_{3,2}\\
                    &+Z_{2,2}Z_{4,3}+Z_{2,3}Z_{4,4}+Z_{4,2}Z_{2,3}+Z_{4,3}Z_{2,4} \\
                 &= Z_{2,2}Z_{2,3} + Z_{2,2}Z_{2,4} + Z_{2,2}Z_{3,2} + Z_{2,2}Z_{4,2} + Z_{2,2}Z_{4,3} - 3Z_{2,2} \\
&+ Z_{2,3}Z_{2,4} + Z_{2,3}Z_{3,3} + Z_{2,3}Z_{4,2} + Z_{2,3}Z_{4,3} + Z_{2,3}Z_{4,4} - 4Z_{2,3} \\
&+ Z_{2,4}Z_{3,4} + Z_{2,4}Z_{4,3} + Z_{2,4}Z_{4,4} - 3Z_{2,4} + Z_{3,2}Z_{3,3} + Z_{3,2}Z_{3,4} + Z_{3,2}Z_{4,2} - 4Z_{3,2} \\
&+ Z_{3,3}Z_{3,4} + Z_{3,3}Z_{4,3} - 2Z_{3,3} + Z_{3,4}Z_{4,4} - 4Z_{3,4} \\
&+ Z_{4,2}Z_{4,3} + Z_{4,2}Z_{4,4} - 3Z_{4,2} + Z_{4,3}Z_{4,4} - 4Z_{4,3} - 3Z_{4,4}  
\end{aligned}
\end{equation}

Finally, we can construct the circuit for the cost hamiltonian. The circuit has four qubits, each represent:
\begin{center}
  \begin{enumerate}
    \item   $Q_1$ represent $x_{2,2}$ 
    \item   $Q_2$ represent $x_{2,3}=1$ 
    \item   $Q_3$ represent $x_{2,4}=1$       
    \item   $Q_4$ represent $x_{3,2}=1$  
    \item   $Q_5$ represent $x_{3,3}=1$      
    \item   $Q_6$ represent $x_{3,4}=1$     
    \item   $Q_7$ represent $x_{4,2}=1$   
    \item    $Q_8$ represent $x_{4,3}=1$    
    \item   $Q_9$ represent $x_{4,4}=1$  
\end{enumerate}  
\end{center}

This format should be more compact while still clearly representing the polynomial.

The final hamiltonian is:

\begin{align*}
 H_C &= Z_1Z_2 + Z_1Z_3 + Z_1Z_4 + Z_1Z_7 + Z_1Z_8 - 3Z_1 + Z_2Z_3 + Z_2Z_5 + Z_2Z_7 \\
 &+ Z_2Z_8 + Z_2Z_9 - 4Z_2 + Z_3Z_6 + Z_3Z_8 + Z_3Z_9 - 3Z_3 + Z_4Z_5 + Z_4Z_6 + Z_4Z_7 \\
 &- 4Z_4 + Z_5Z_6 + Z_5Z_8 - 2Z_5 + Z_6Z_9 - 4Z_6 + Z_7Z_8 + Z_7Z_9 - 3Z_7 + Z_8Z_9 - 4Z_8 - 3Z_9   
\end{align*}

Since the correct result \footnote{The benefit to use a cycle to be the example is that there are only two solution, clockwise and counterclockwise, and thus we can easily check the correctness of our hamiltonian} must be either $(1,0,0,0,1,0,0,0,1)$, which represent the hamiltonian cycle $(1 \rightarrow 2\rightarrow3\rightarrow \rightarrow 4 \rightarrow 1 )$ or $(0,0,1,0,1,0,1,0,0)$, which represent the hamiltonian cycle $(1 \rightarrow 4\rightarrow 3\rightarrow 2 \rightarrow 1 )$. The ground state for this solution must in dirac notation be either $\ket{100010001}$ or $\ket{001010100}$. We can check the correctness by calculating the eigen value and eigen energies of the above equation:

\begin{lstlisting}
 import numpy as np
from scipy.linalg import eigh
from functools import reduce

# Define the Pauli Z matrix
pauli_z = np.array([[1, 0], [0, -1]])

# Function to create a matrix representation of Z_k gate on k-th qubit
def z_k_matrix(k, total_qubits):
    I = np.eye(2)  # Identity matrix
    matrices = [pauli_z if i == k else I for i in range(total_qubits)]
    return reduce(np.kron, matrices)

# Total number of qubits
total_qubits = 9

# Construct the Hamiltonian matrix
H = np.zeros((2**total_qubits, 2**total_qubits))

# Define the terms of the Hamiltonian (coefficients and qubit indices)
terms = [
    (1, [0, 1]), (1, [0, 2]), (1, [0, 3]), (1, [0, 6]), (1, [0, 7]),
    (-3, [0]), (1, [1, 2]), (1, [1, 4]), (1, [1, 6]), (1, [1, 7]),
    (1, [1, 8]), (-4, [1]), (1, [2, 5]), (1, [2, 7]), (1, [2, 8]),
    (-3, [2]), (1, [3, 4]), (1, [3, 5]), (1, [3, 6]), (-4, [3]),
    (1, [4, 5]), (1, [4, 7]), (-2, [4]), (1, [5, 8]), (-4, [5]),
    (1, [6, 7]), (1, [6, 8]), (-3, [6]), (1, [7, 8]), (-4, [7]),
    (-3, [8])
]

# Add each term to the Hamiltonian
for coeff, qubits in terms:
    if len(qubits) == 1:
        H += coeff * z_k_matrix(qubits[0], total_qubits)
    else:
        term = z_k_matrix(qubits[0], total_qubits)
        for qubit in qubits[1:]:
            term = np.dot(term, z_k_matrix(qubit, total_qubits))
        H += coeff * term

# Calculate eigenvalues and eigenvectors
eigenvalues, eigenvectors = eigh(H)

# Extract the five lowest eigenvalues and their corresponding eigenstates
lowest_five_eigenvalues = eigenvalues[:5]
lowest_five_eigenstates = eigenvectors[:, :5]

# Convert the eigenstates to Dirac notation
lowest_five_eigenstates_dirac = []
for i in range(5):
    max_amplitude_index = np.argmax(np.abs(lowest_five_eigenstates[:, i]))
    dirac_state = "|{0:09b}>".format(max_amplitude_index)
    lowest_five_eigenstates_dirac.append(dirac_state)

# Display the results
print("Lowest Five Eigenvalues:", lowest_five_eigenvalues)
print("Corresponding Eigenstates in Dirac Notation:", 
    lowest_five_eigenstates_dirac)
\end{lstlisting}

\begin{figure}[h!]
    \centering
    \includegraphics[width=1\linewidth]{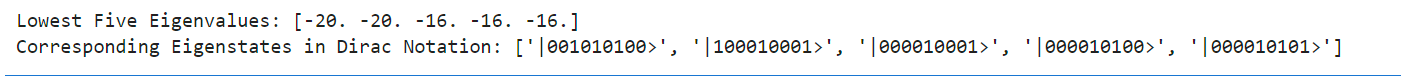}
    \caption{The output has proved the correctness of our hamiltonian,  because the eigen states with the lowest energies are $\ket{001010100}$ and $\ket{100010001}$, which are the exact solution for the hamiltonian path of a square.}
    \label{fig:Eigen}
\end{figure}

\begin{figure}[h!]
    \centering
    \includegraphics[width=1\linewidth]{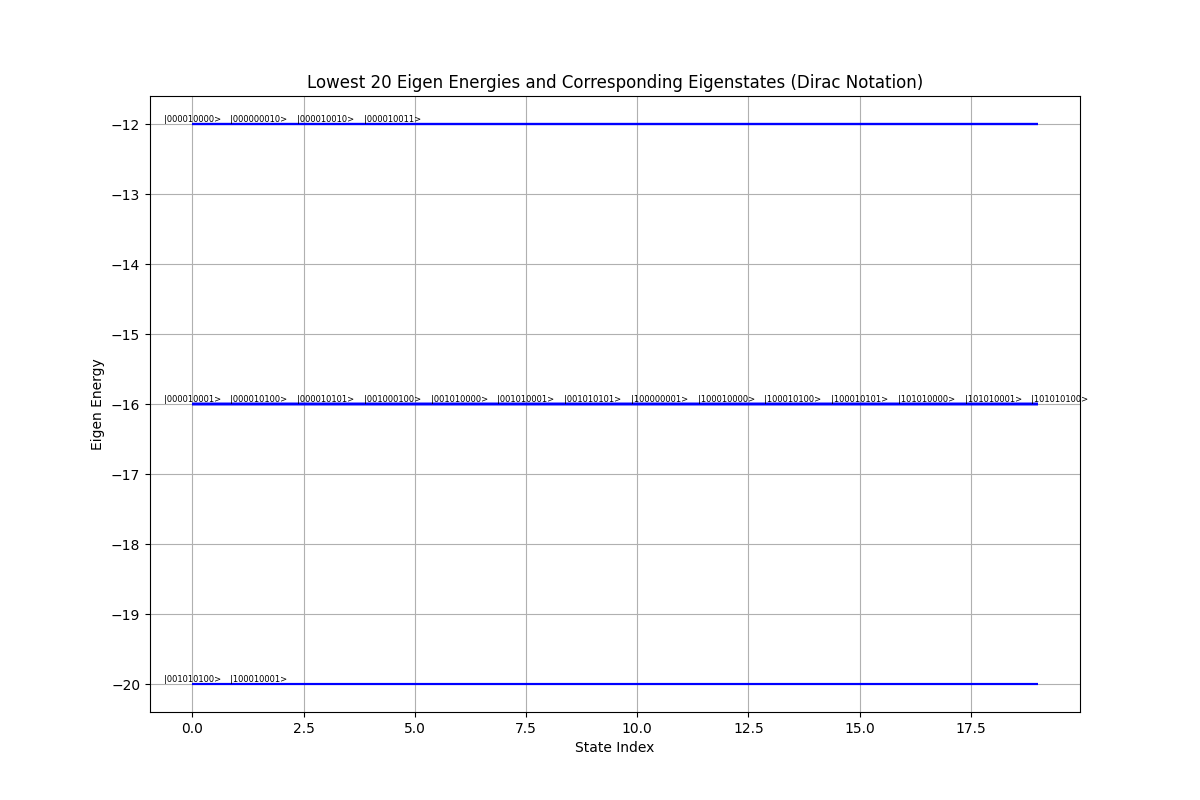}
    \caption{Plot the energy spectrum of the cost hamiltonian. The minimum energy is -20, and the eigen states are $\ket{001010100}$ and  $\ket{100010001}$. The energy spectrum justify the design of our hamiltonian for hamiltonian cycle because there is a huge energy gap between the solution we want and the other solutions.}
    \label{fig:FourEigen}
\end{figure}

\section{Simulation using Qiskit}

\subsection{How do I scale the problem up}
The most difficult part in implementation is \textbf{how to embed an arbitrary graph automatically to cost hamiltonian in QAOA?}.
I write a function, the input is the graph, the output is the the parameter for the hamiltonian, in a python dictionary.

\begin{lstlisting}

from sympy import symbols, Sum, IndexedBase, simplify
from sympy.abc import n, v, j, u

# Define symbolic variables
x = IndexedBase('x')

# Function to represent the vertex uniqueness term of the Hamiltonian
def vertex_uniqueness_term(n):
    return Sum((1 - Sum(x[v, j], (j, 2, n)))**2, (v, 2, n))

# Function to represent the edge uniqueness term of the Hamiltonian
def edge_uniqueness_term(n):
    return Sum((1 - Sum(x[v, j], (v, 2, n)))**2, (j, 2, n))

# Function to represent the edge validity term 
# of the Hamiltonian for a given graph
def edge_validity_term(graph, n):
    validity_term = 0
    for u in range(2,n+1):
        edge=(u,1)
        if not edge in graph:  
            validity_term +=x[u,n]
        for v in range(2,n+1):
            u, v = edge
            if not edge in graph:
                validity_term += Sum(x[u, j] * x[v, j+1], (j, 1, n-1))                
    return validity_term

# Combine the terms to form the complete Hamiltonian
def hamiltonian(graph, n):
    H = vertex_uniqueness_term(n) + 
        edge_uniqueness_term(n) + edge_validity_term(graph, n)
    return simplify(H)


def apply_substitution_to_hamiltonian(H, n):
    Z = IndexedBase('Z')
    H_substituted = H
    for v in range(2, n+1):
        for j in range(2, n+1):
            z_index = (v-2)*(n-1) + j-1  # Corrected index calculation
            if z_index > 0:
                H_substituted = 
                    H_substituted.subs(x[v, j], 1/2 * (1 - Z[z_index]))
            else:
                H_substituted = H_substituted.subs(x[v, j], 0)
    return simplify(H_substituted)


def expand_and_simplify_hamiltonian(H,n):
    Z = IndexedBase('Z')
    H_expanded = H.expand()
    # Apply the simplification rule Z_k^2 = I
    for k in range(1, (n-1)**2+1):  # Assuming up to 8 qubits for this example
        H_expanded = H_expanded.subs(Z[k]**2, 0)
    return simplify(H_expanded)


def hamiltonian_to_string_list(H, n):
    """
    Convert the expanded Hamiltonian to a 
    list of strings with corresponding coefficients.
    Each string represents a term in the Hamiltonian,
    with 'Z' at positions corresponding to qubits involved in the term.
    For example, 'ZZI' represents Z_1 Z_2.
    
    :param H: The expanded Hamiltonian expression
    :param n: Number of qubits
    :return: List of tuples (string, coefficient)
    """
    Z = IndexedBase('Z')
    terms = []
    
    # Iterate over each term in the Hamiltonian expression
    for term in H.as_ordered_terms():
        # Initialize a string with 'I's for each qubit
        term_string = ['I'] * n
        coeff = H.coeff(term)  # Extract the coefficient of the term
        
        # Check for the presence of Z operators in the term
        for k in range(1, n+1):
            if term.has(Z[k]):
                term_string[k-1] = 'Z'
        
        # Join the term string and append it with its coefficient to the list
        terms.append((''.join(term_string), coeff))
    
    return terms

n=3
# Example: Hamiltonian for a triangle graph
triangle_graph = [(1, 2), (2, 3), (3, 1), (2, 1), (3, 2), (3, 1)]
H_triangle = hamiltonian(triangle_graph, n)
print(f"Polynomial H is {H_triangle}")
# Apply the substitution rule to the
# Hamiltonian for a triangle graph (n = 3)
H_triangle_substituted = apply_substitution_to_hamiltonian(H_triangle, n)
print(f"After Substitute wo Z is {H_triangle_substituted}")
H_final=expand_and_simplify_hamiltonian(H_triangle_substituted,n)
print(f"After simplification  {H_final}")
# Convert the final Hamiltonian for the 
# triangle graph to string list representation
hamiltonian_string_list = hamiltonian_to_string_list(H_final, (n-1)**2)
print(f"Final result  {hamiltonian_string_list}")
\end{lstlisting}

\begin{figure}[h!]
    \centering
    \includegraphics[width=1\linewidth]{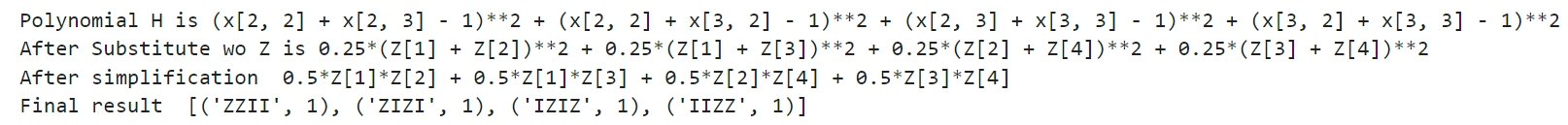}
    \caption{The output after I execute the above code, which calculate the final cost hamiltonian of a triangle according to \autoref{eq:HCircle}. The output is just the same as what I've calculated by hand.}
    \label{fig:Hccalc}
\end{figure}

The output in \autoref{fig:Hccalc} is exactly the form of hamiltonian that I calculated by hand based on \autoref{eq:HCircle}. With the code above, I can generate the cost hamiltonian for hamiltonian cycle problem for an arbitrary graph. Qiskit has already implement the next step to compile the dictionary to the circuit.

\subsection{Simulation of QAOA for Hamiltonian cycle of a triangle.}

\begin{figure}[h!]
    \centering
    \includegraphics[width=1\linewidth]{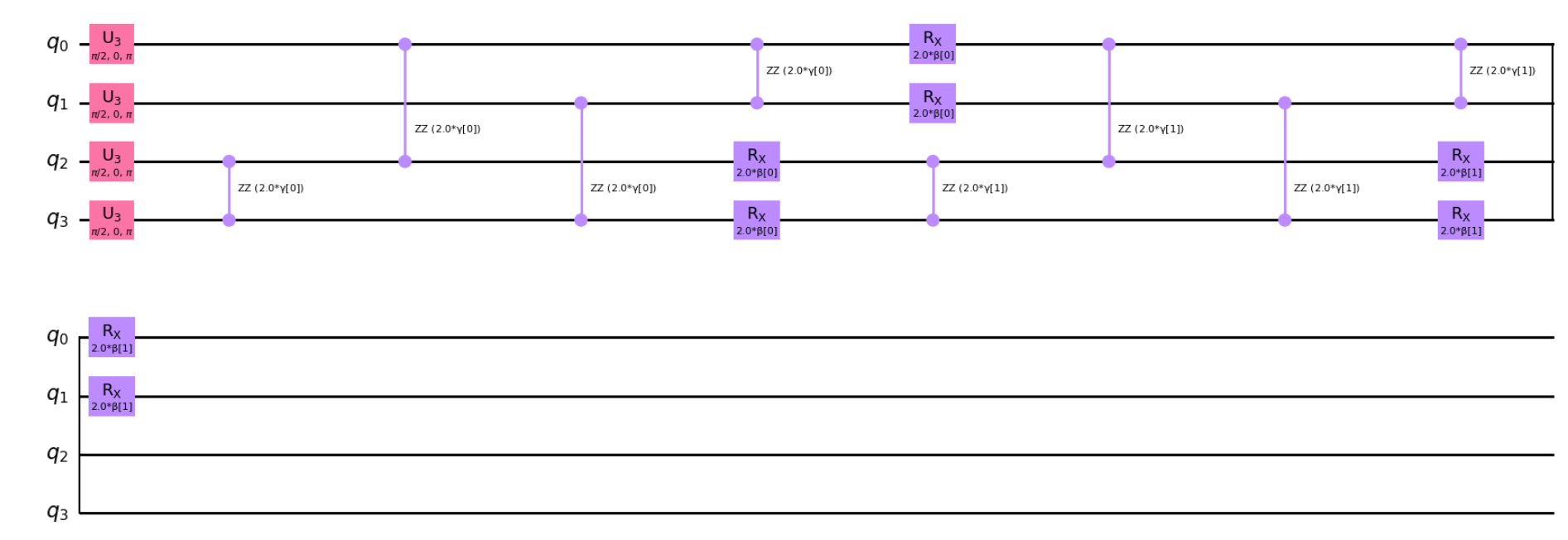}
    \caption{The QAOA circuit for hamiltonian cycle  on a triangle that I used in the experiment. There are four qubits and 2 total repetitions of mixer and cost circuit. }
    \label{fig:HcWholeCircuit}
\end{figure}

First, I run the simulation of QAOA algorithm on the simplest case: When the graph is a triangle! The benefit of doing this is that the circuit is really simple and it is easy for us to test the correctness. The compiled circuit, when the repetition number is two, is shown in \autoref{fig:HcWholeCircuit}. After a row of hadamard gate, there are four pairs of $ZZ$ gate, as defined and calculated in \autoref{eq:HamiltonianTriangle}, which is the cost circuit. The mixer part, is chosen as a row of $R_x$ gate, with the same rotation angle. Finally, we measure the result and get the output. Since our goal is to minimize the energy of the cost hamiltonian with respect to the output state, we have to to calculate the cost value\footnote{In classcal simulation, the cost is easy to calculate, because we only need to the the matrix vector calculation:$\bra{\psi} H\ket{\psi}$. However, in a real quantum computer, to get $\bra{\psi} H\ket{\psi}$ is much harder. Generally speaking, we have to divide $H$ into different Pauli Gate, and measure the expectation of each pauli gate by sampling. Finally, add the energy of each pauli gate.} and use a classical optimizer to minimize the energy. We use the  "COBYLA" method of scipy.optimize to to the optimization.

\begin{figure}[h!]
    \centering
    \includegraphics[width=1\linewidth]{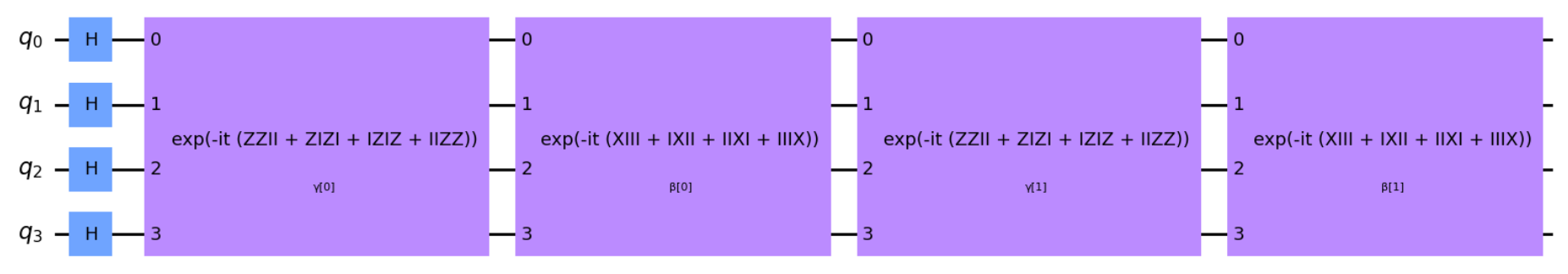}
    \caption{The QAOA circuit block demonstration. There are four block of hamiltonian, each with its own parameter.}
    \label{fig:HcTriangle}
\end{figure}

\begin{figure}[h!]
    \centering
    \includegraphics[width=1\linewidth]{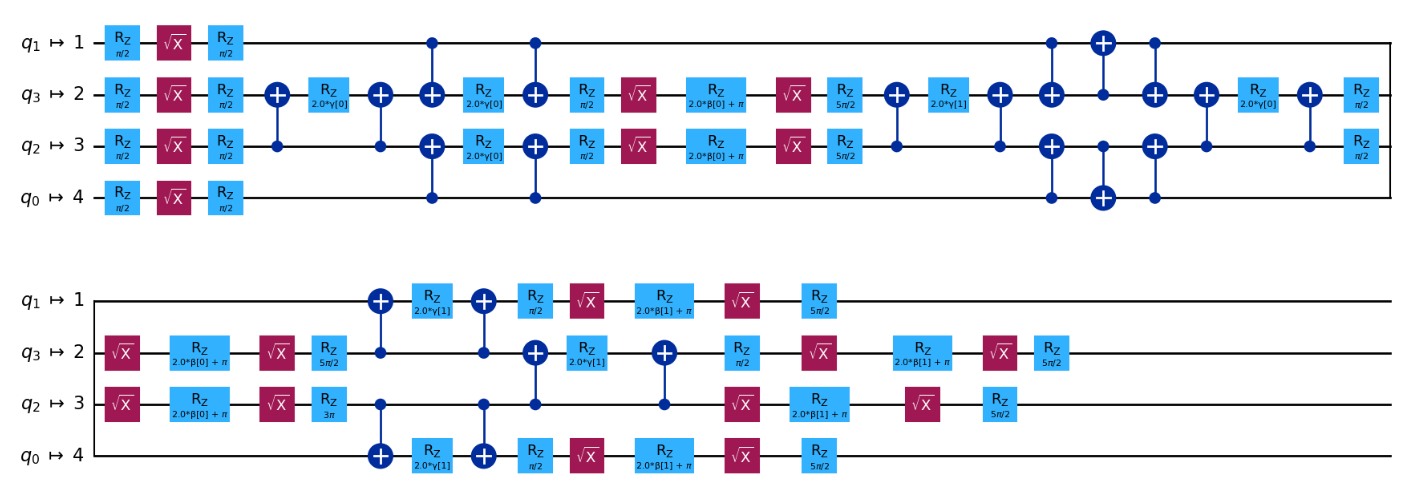}
    \caption{The qubit mapping to the FakeManilaV2 after transpilation.}
    \label{fig:triangleMapping}
\end{figure}

\begin{figure}[h!]
    \centering
    \includegraphics[width=1\linewidth]{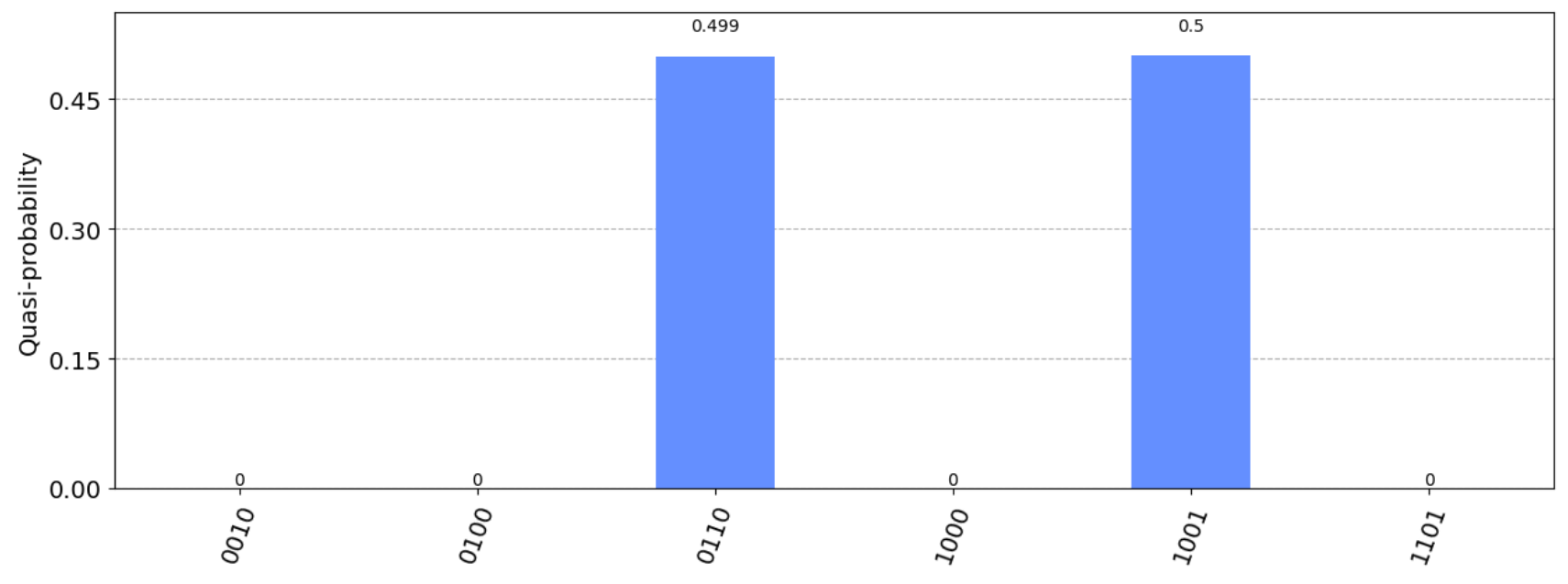}
    \caption{The result of QAOA solving hamiltonian cycle. The final probability of measurement, after optimizing the parameters. The result match with our previous expectation. }
    \label{fig:Circle3NoNoise}
\end{figure}

\begin{figure}[h!]
    \centering
    \includegraphics[width=1\linewidth]{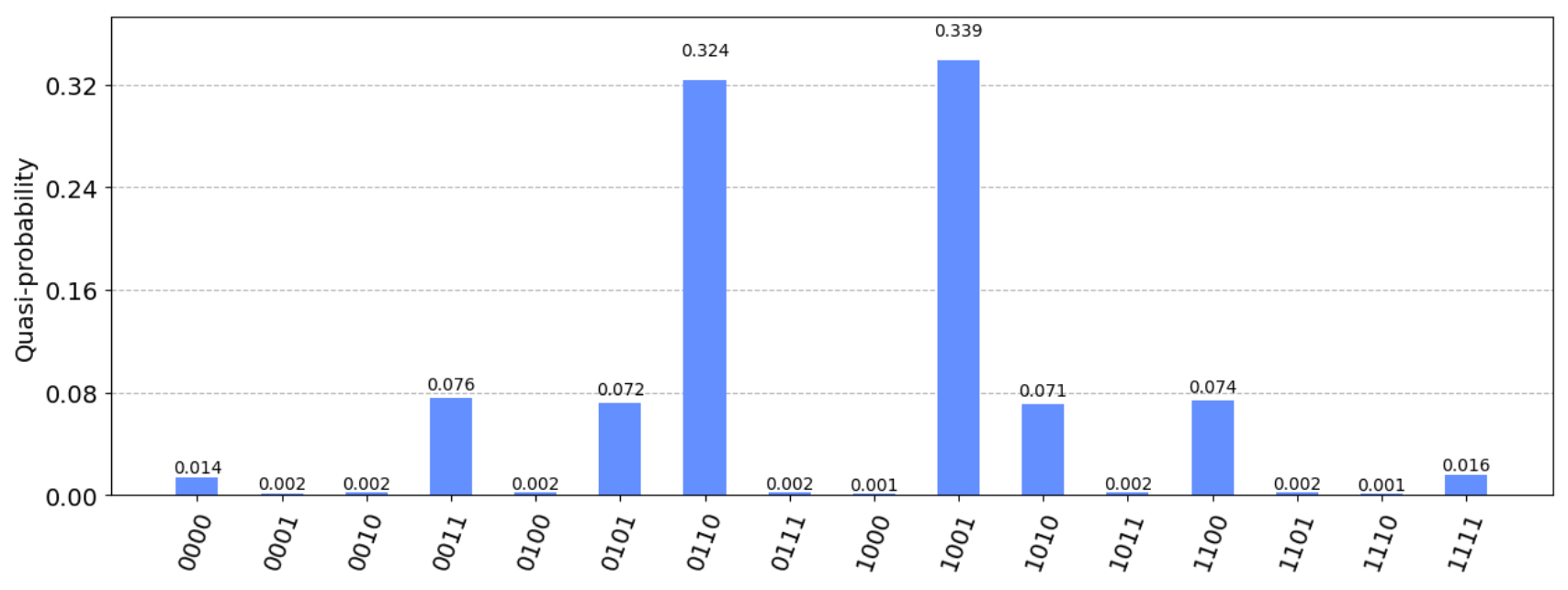}
    \caption{The result of running QAOA for hamiltonian cycle on a triangle on a fake noisy quantum chip. We use FakeManilaV2 to the simulation. Although the success probability drp a little bit, we will still get correct result, with hight probability.}
    \label{fig:FakeManila}
\end{figure}

\begin{figure}[h!]
    \centering
    \includegraphics[width=1\linewidth]{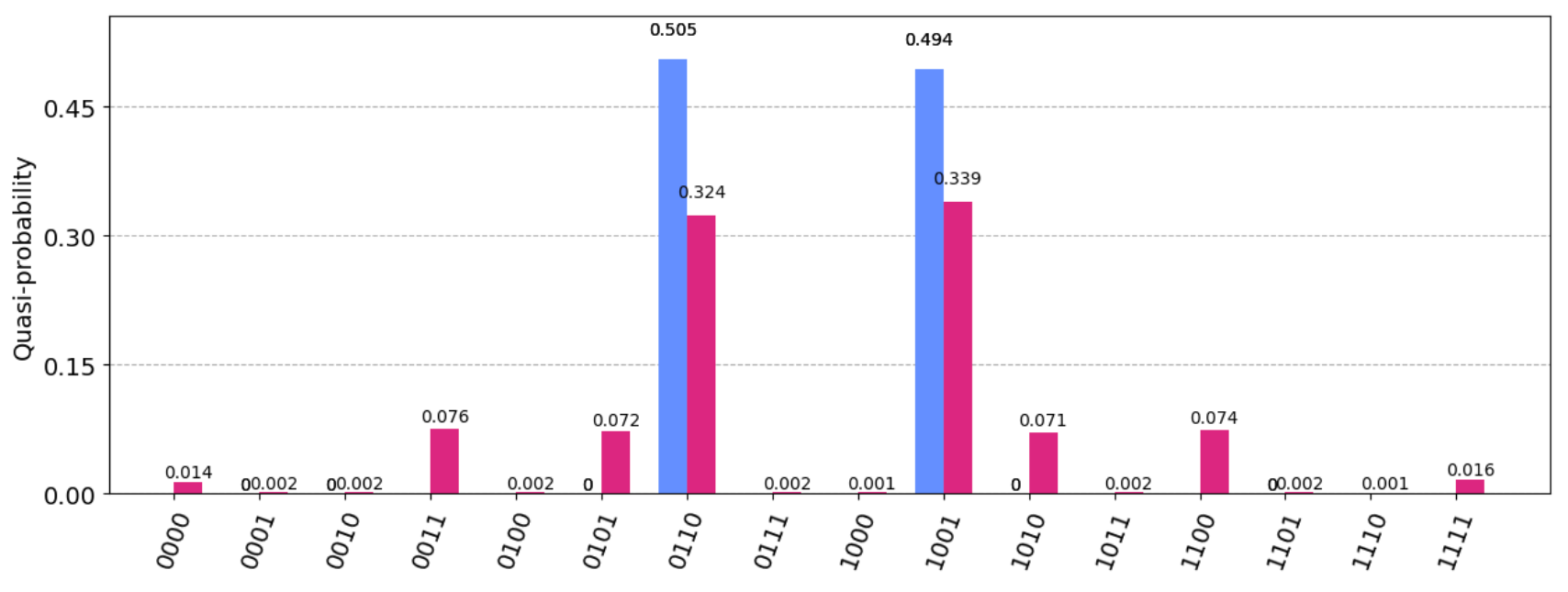}
    \caption{Compare the noiseless result in \autoref{fig:Circle3NoNoise} result and the noisy result \autoref{fig:FakeManila} of a triangle case. }
    \label{fig:FakeManilaCompare}
\end{figure}

\FloatBarrier

\subsection{Simulation of QAOA for Hamiltonian cycle of a Square}

\begin{figure}[h!]
    \centering
    \includegraphics[width=1\linewidth]{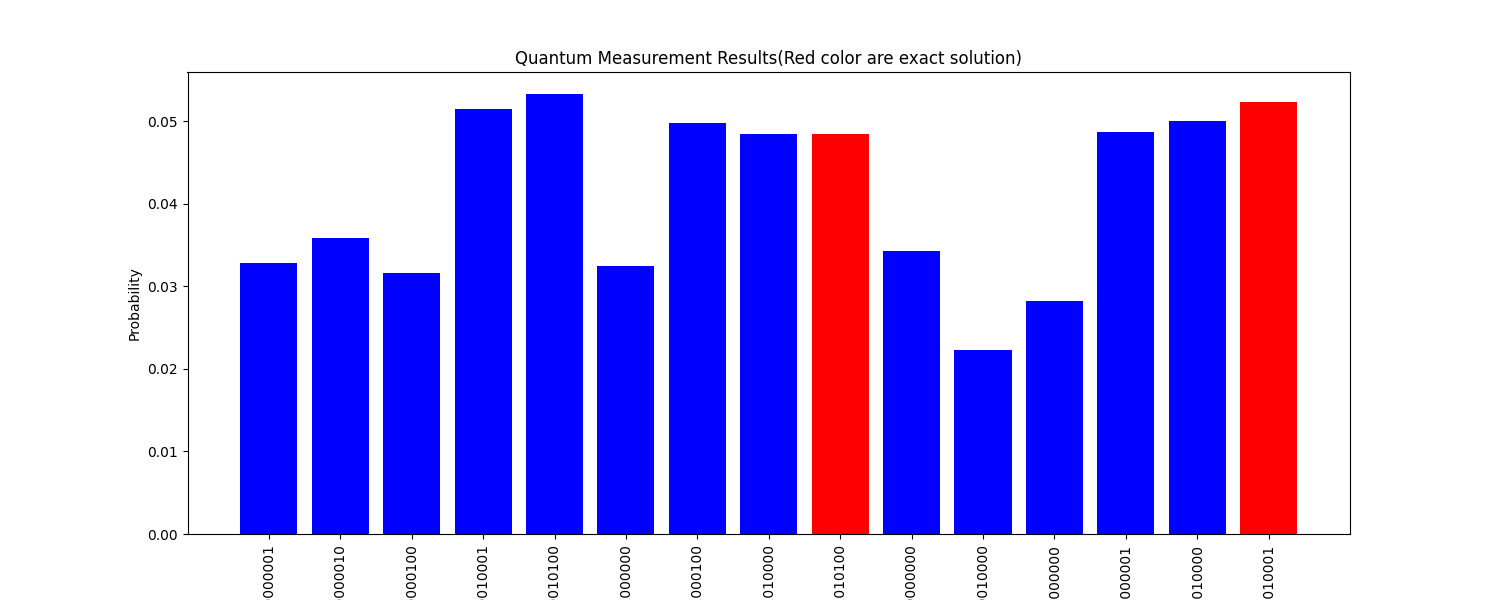}
    \caption{The experimental result of QAOA for hamiltonian cycle without noise when the graph is a square. I use red color to highlight the two exact solution $\ket{001010100}$ and  $\ket{100010001}$.}
    \label{fig:squareNoNoise}
\end{figure}

\begin{figure}[h!]
    \centering
    \includegraphics[width=1\linewidth]{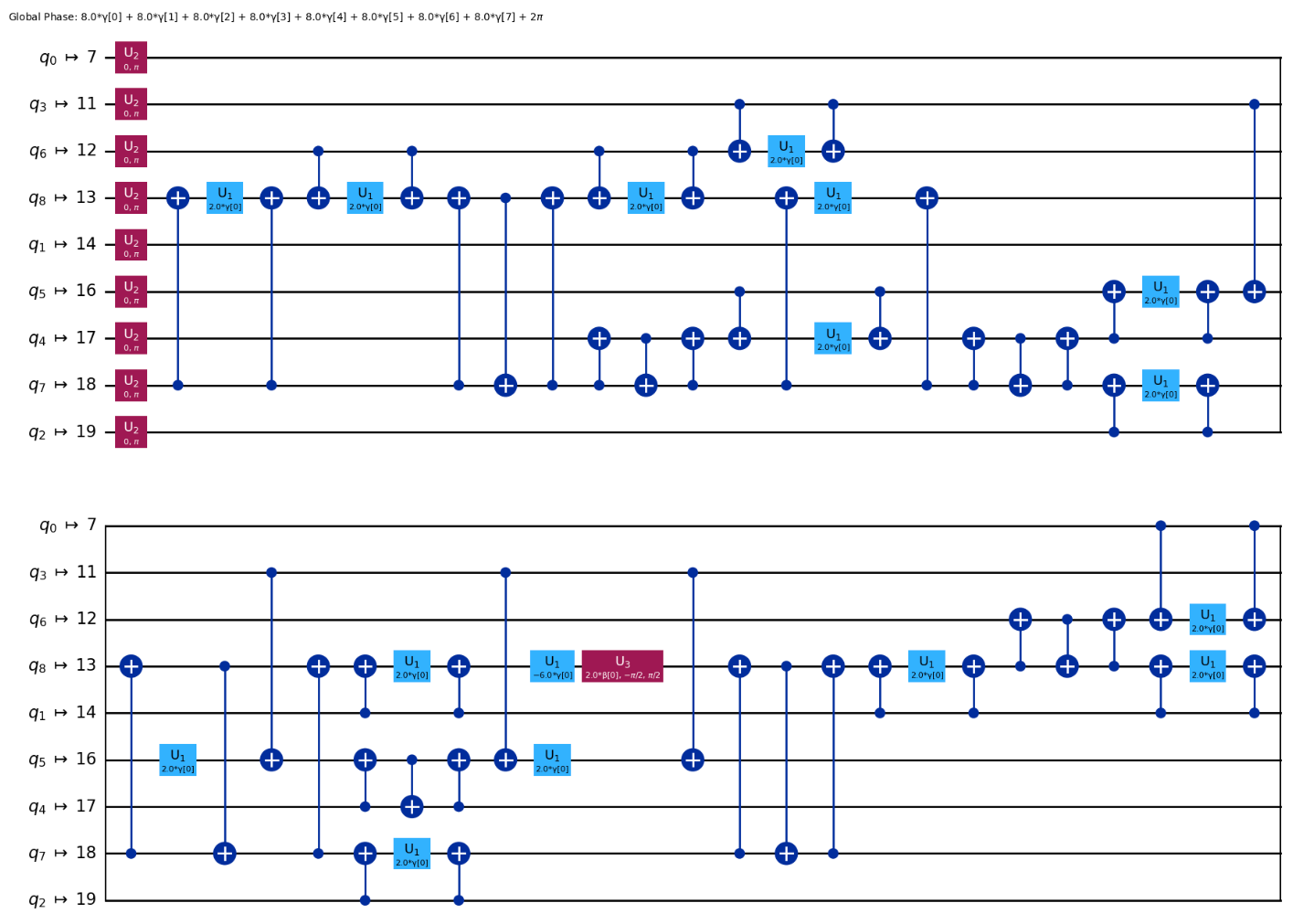}
    \caption{The qubit mapping to the FakeAlmadenV2 that we use for solving the hamiltonian path problem on a square. }
    \label{fig:squareMapping}
\end{figure}

\begin{figure}[h!]
    \centering
    \includegraphics[width=1\linewidth]{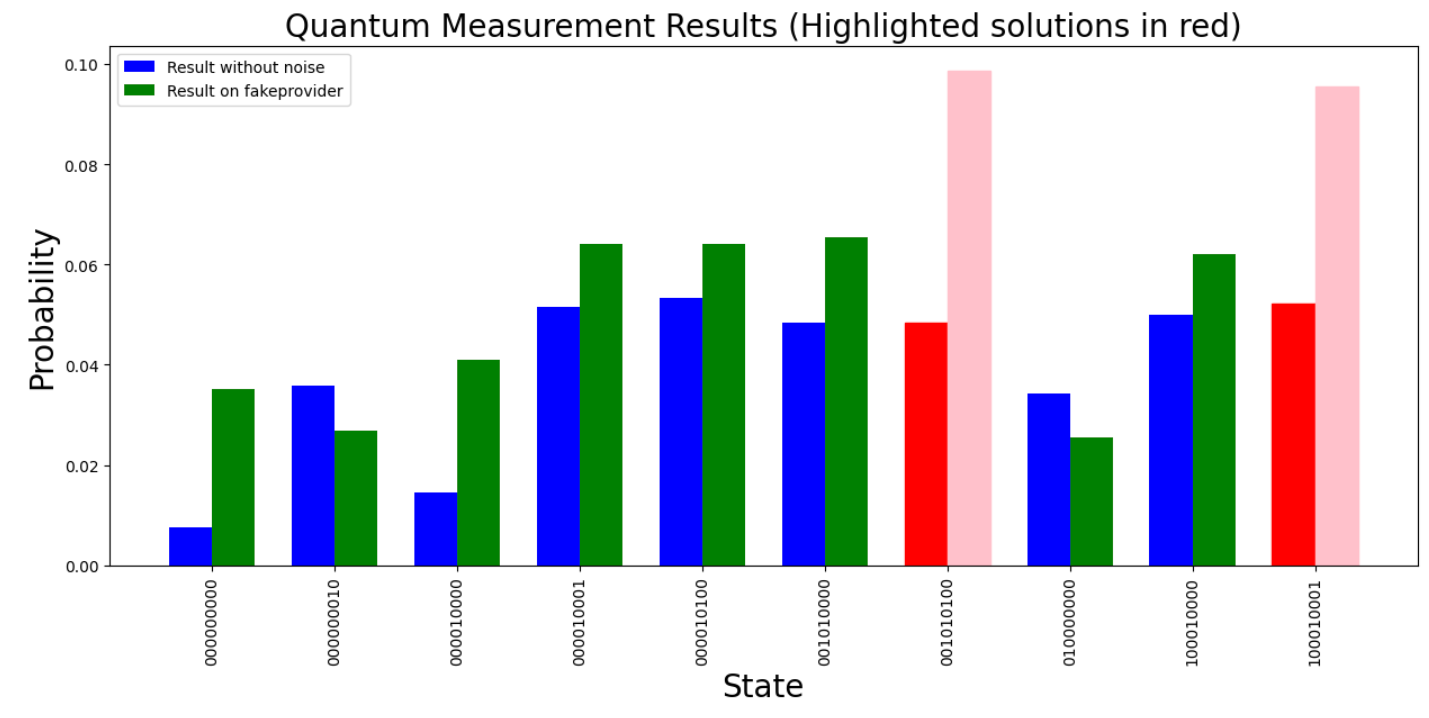}
    \caption{The experimental result of QAOA for hamiltonian cycle with noise. Amazingly, the result actually is much better than without noise! }
    \label{fig:squareNoise}
\end{figure}

\begin{figure}[h!]
    \centering
    \includegraphics[width=1\linewidth]{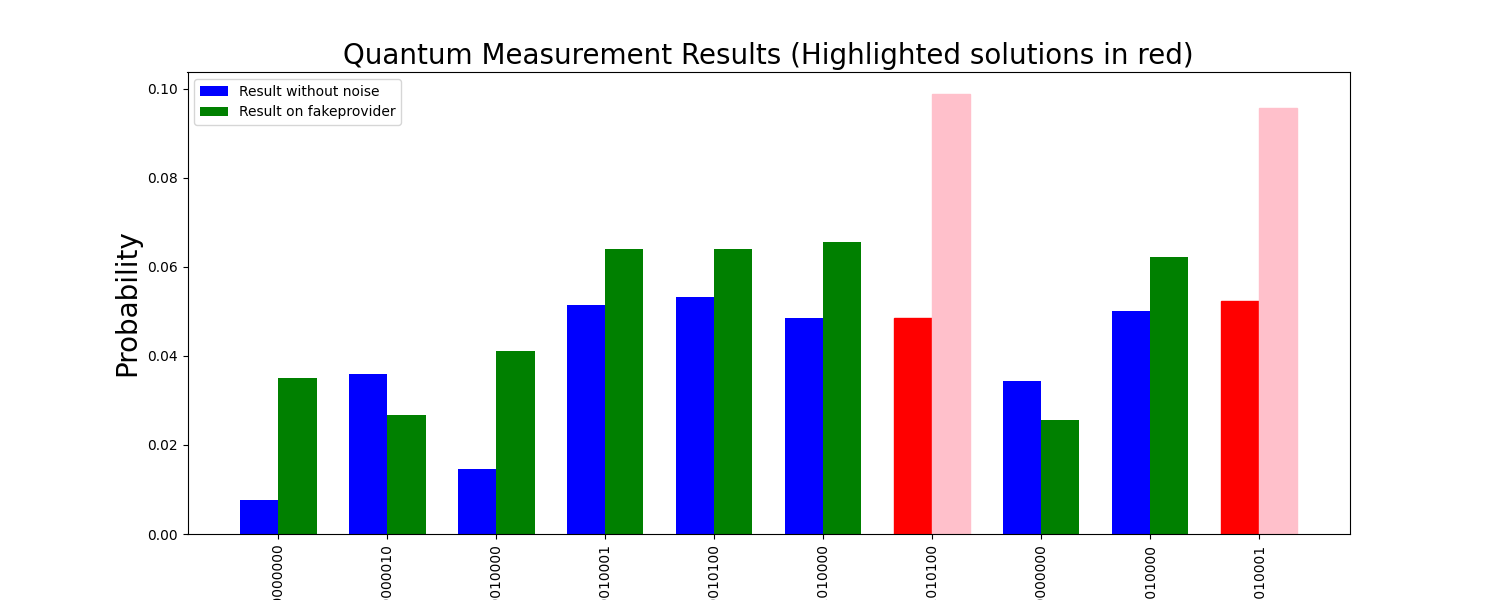}
    \caption{Compare the result of noiseless simulation in \autoref{fig:squareNoNoise} and noisy simulation \autoref{fig:squareNoise} together. The blue color denotes the result of simulation on a noiseless simulator and the green color denotes the result of a noisy simulator. I highlight the correct result,$\ket{001010100}$ and $\ket{100010001}$ in red for the noiseless result and in pink for the noisy result. The final probability of getting the accurate result is much higher in the FakeAlmadenV2 simulator! }
    \label{fig:squareCompare}
\end{figure}

\subsection{Result of different Mixers}

The choice of mixer can be essential in the training and optimization process of QAOA. In the previous, I used the default $R_x$ mixer. In this section, I also use $R_y$ mixer to run the QAOA and compare the new result with the previous one.

\begin{figure}[h!]
    \centering
    \includegraphics[width=1\linewidth]{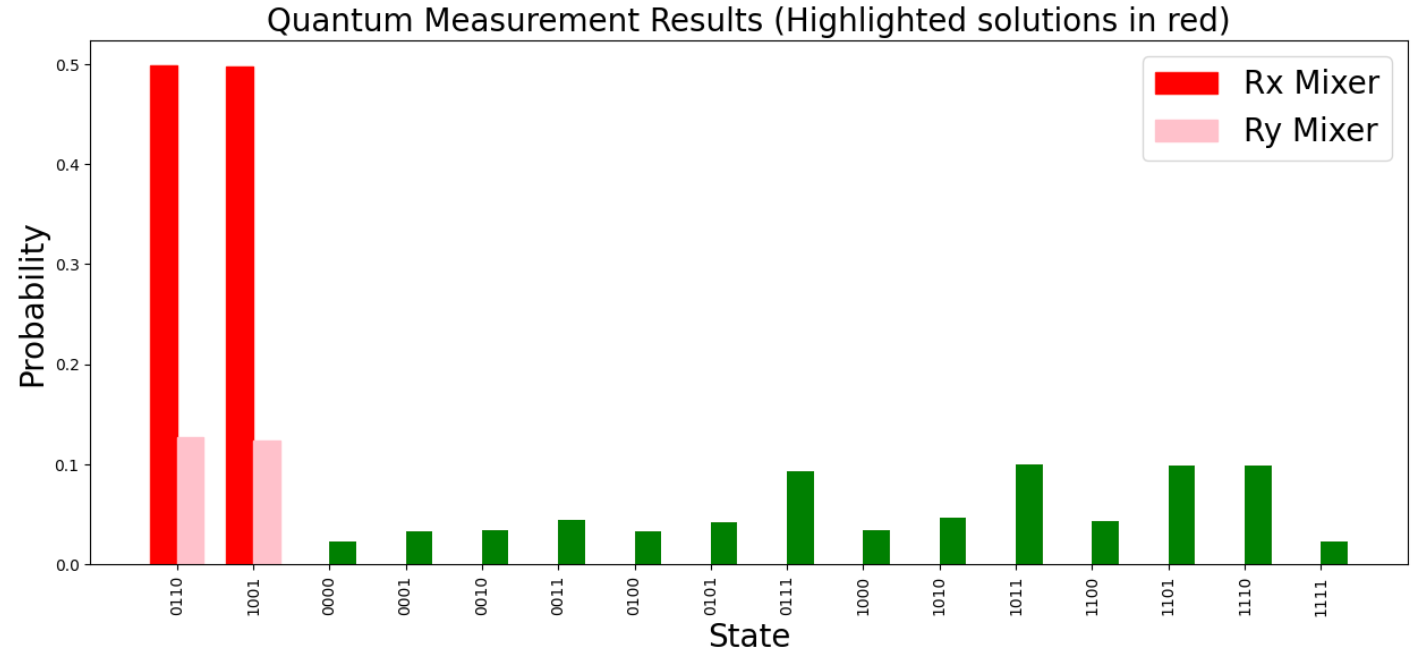}
    \caption{Compare the QAOA for hamiltonian cycle on a triangle result between two different mixers. The two outstanding red bars represent the measurement of the correct solution by circuit with $R_x$ mixer, while two tiny pinks bars are the result of simulation by circuit with $R_y$ mixer.}
    \label{fig:mixer3}
\end{figure}

\begin{figure}[h!]
    \centering
    \includegraphics[width=1\linewidth]{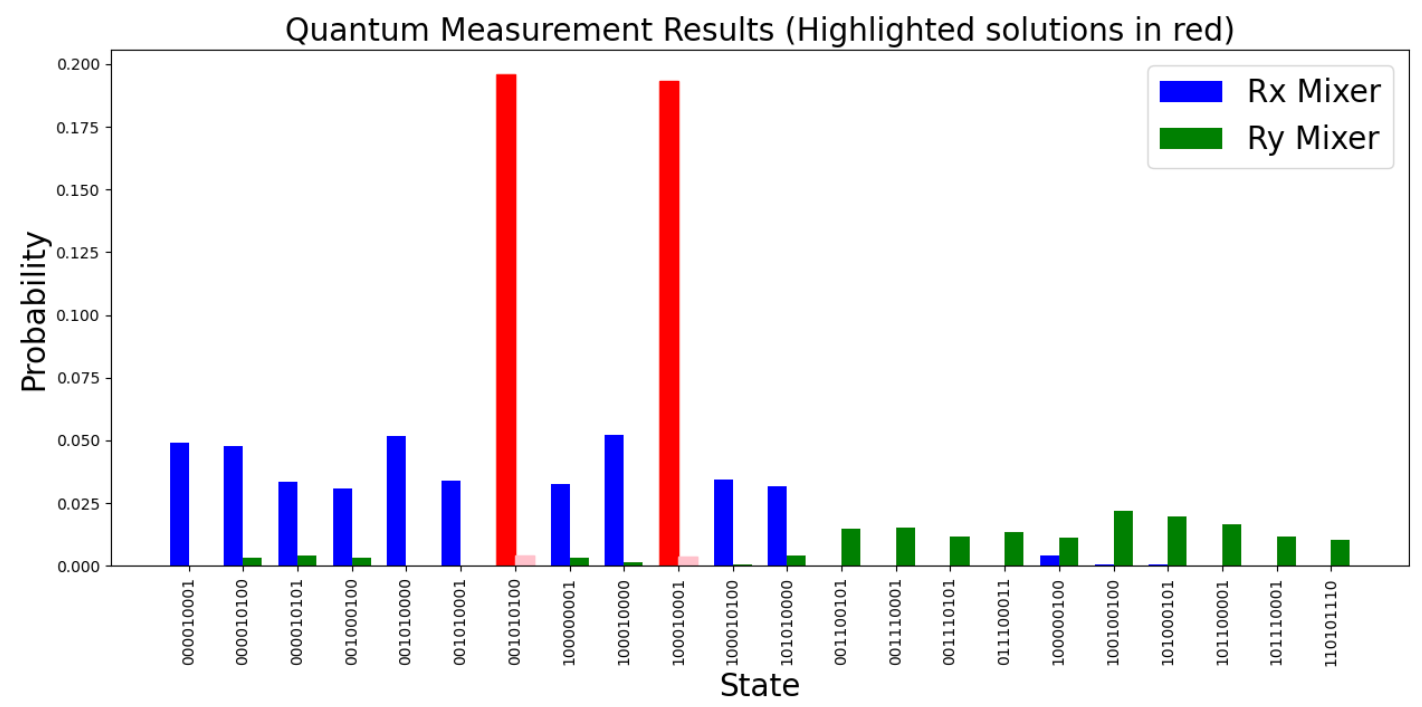}
    \caption{Compare the QAOA for hamiltonian cycle on a square result between two different mixers. The two outstanding red bars represent the measurement of the correct solution by circuit with $R_x$ mixer, while two tiny pinks bars are the result of simulation by circuit with $R_y$ mixer.}
    \label{fig:mixer4}
\end{figure}

In \autoref{fig:mixer3}, we run the same noiseless simulation for hamiltonian cycle on a triangle, and compare the result of $R_x$ mixer and $R_y$ mixer. I also simulate for the same comparison when the graph is a square, which is plotted in \autoref{fig:mixer4}. From the result in \autoref{fig:mixer3} and \autoref{fig:mixer4}, it's clear that $R_x$ mixer is much better than $R_y$ mixer.

\FloatBarrier

\section{Code for my simulation}

\begin{lstlisting}
# General imports
import numpy as np

# Pre-defined ansatz circuit, operator class and visualization tools
from qiskit.circuit.library import QAOAAnsatz
from qiskit.quantum_info import SparsePauliOp
from qiskit.visualization import plot_distribution
from qiskit.providers.fake_provider import FakeManilaV2
# Qiskit Runtime
from qiskit_ibm_runtime import QiskitRuntimeService
from qiskit_ibm_runtime import Estimator, Sampler, Session, Options
from qiskit.providers.fake_provider import FakeManilaV2
# SciPy minimizer routine
from scipy.optimize import minimize
from qiskit.primitives import Estimator, Sampler
options = Options()
options.transpilation.skip_transpilation = False
options.execution.shots = 10000
estimator = Estimator(options={"shots": int(1e4)})
sampler = Sampler(options={"shots": int(1e4)})

from sympy import symbols, Sum, IndexedBase, simplify
from sympy.abc import n, v, j, u

# Define symbolic variables
x = IndexedBase('x')

# Function to represent the vertex uniqueness term of the Hamiltonian
def vertex_uniqueness_term(n):
    return Sum((1 - Sum(x[v, j], (j, 2, n)))**2, (v, 2, n))

# Function to represent the edge uniqueness term of the Hamiltonian
def edge_uniqueness_term(n):
    return Sum((1 - Sum(x[v, j], (v, 2, n)))**2, (j, 2, n))

# Function to represent the edge validity 
# term of the Hamiltonian for a given graph
def edge_validity_term(graph, n):
    validity_term = 0
    for u in range(2,n+1):
        edge=(u,1)
        if not edge in graph:  
            validity_term +=x[u,n]
        for v in range(2,n+1):
            u, v = edge
            if not edge in graph:
                validity_term += Sum(x[u, j] * x[v, j+1], (j, 1, n-1))                
    return validity_term

# Combine the terms to form the complete Hamiltonian
def hamiltonian(graph, n):
    H = 1.5*vertex_uniqueness_term(n) +
        edge_uniqueness_term(n) + edge_validity_term(graph, n)
    return simplify(H)


def apply_substitution_to_hamiltonian(H, n):
    Z = IndexedBase('Z')
    H_substituted = H
    for v in range(2, n+1):
        for j in range(2, n+1):
            z_index = (v-2)*(n-1) + j-1  # Corrected index calculation
            if z_index > 0:
                H_substituted = 
                    H_substituted.subs(x[v, j], 1/2 * (1 - Z[z_index]))
            else:
                H_substituted = H_substituted.subs(x[v, j], 0)
    return simplify(H_substituted)


def expand_and_simplify_hamiltonian(H,n):
    Z = IndexedBase('Z')
    H_expanded = H.expand()
    # Apply the simplification rule Z_k^2 = I
    for k in range(1, (n-1)**2+1): 
        # Assuming up to 8 qubits for this example
        H_expanded = H_expanded.subs(Z[k]**2, 0)
    return simplify(H_expanded)


def hamiltonian_to_string_list(H, n):
    """
    Convert the expanded Hamiltonian to a list of 
    strings with corresponding coefficients.
    Each string represents a term in the Hamiltonian,
    with 'Z' at positions corresponding to qubits involved in the term.
    For example, 'ZZI' represents Z_1 Z_2.
    
    :param H: The expanded Hamiltonian expression
    :param n: Number of qubits
    :return: List of tuples (string, coefficient)
    """
    Z = IndexedBase('Z')
    terms = []
    
    # Iterate over each term in the Hamiltonian expression
    for term in H.as_ordered_terms():
        # Initialize a string with 'I's for each qubit
        term_string = ['I'] * n
        coeff = H.coeff(term)  # Extract the coefficient of the term
        findZ=False
        # Check for the presence of Z operators in the term
        for k in range(1, n+1):
            if term.has(Z[k]):
                findZ=True
                term_string[k-1] = 'Z'
        if not findZ:
            continue
        # Join the term string and append it with its coefficient to the list
        terms.append((''.join(term_string), coeff))
    
    return terms

n=4
# Example: Hamiltonian for a triangle graph
triangle_graph = [(1, 2), (2, 3), (3, 4),(4,1), (2, 1), (3, 2),(4,3),(1,4)]
H_triangle = hamiltonian(triangle_graph, n)
print(f"Polynomial H is {H_triangle}")
# Apply the substitution rule to the Hamiltonian for a triangle graph (n = 3)
H_triangle_substituted = apply_substitution_to_hamiltonian(H_triangle, n)
print(f"After Substitute wo Z is {H_triangle_substituted}")
H_final=expand_and_simplify_hamiltonian(H_triangle_substituted,n)
print(f"After simplification  {H_final}")
# Convert the final Hamiltonian for the triangle 
# graph to string list representation
hamiltonian_string_list = hamiltonian_to_string_list(H_final, (n-1)**2)
print(f"Final result  {hamiltonian_string_list}")

from qiskit.providers.fake_provider import FakeAlmadenV2
# Get a fake backend from the fake provider
backend = FakeAlmadenV2()

from qiskit.transpiler import PassManager
from qiskit.transpiler.preset_passmanagers import generate_preset_pass_manager
from qiskit_ibm_runtime.transpiler.passes.scheduling import (
    ALAPScheduleAnalysis,
    PadDynamicalDecoupling,
)
from qiskit.circuit.library import XGate

target = backend.target
pm = generate_preset_pass_manager(target=target, optimization_level=3)


ansatz_ibm = pm.run(ansatz)


hamiltonian_string_list = [
    ('ZZIIIIIII', 1), ('ZIZIIIIII', 1), ('ZIIZIIIII', 1), ('ZIIIIIZII', 1),
    ('ZIIIIIIZI', 1), ('ZIIIIIIII', -3), ('IZZIIIIII', 1), ('IZIIZIIII', 1),
    ('IZIIIIZII', 1), ('IZIIIIIZI', 1), ('IZIIIIIIZ', 1), ('IZIIIIIII', -4),
    ('IIZIIZIII', 1), ('IIZIIIIZI', 1), ('IIZIIIIIZ', 1), ('IIZIIIIII', -3),
    ('IIIZZIIII', 1), ('IIIZIZIII', 1), ('IIIZIIZII', 1), ('IIIZIIIII', -4),
    ('IIIIZZIII', 1), ('IIIIZIIZI', 1), ('IIIIZIIII', -2), ('IIIIIZIIZ', 1),
    ('IIIIIZIII', -4), ('IIIIIIZZI', 1), ('IIIIIIZIZ', 1), ('IIIIIIZII', -3),
    ('IIIIIIIZZ', 1), ('IIIIIIIZI', -4), ('IIIIIIIIZ', -3)
]


# Problem to Hamiltonian operator
hamiltonian = SparsePauliOp.from_list(hamiltonian_string_list)
# QAOA ansatz circuit
ansatz = QAOAAnsatz(hamiltonian, reps=8)

ansatz.decompose(reps=8).draw(output="mpl", style="iqp")

def cost_func(params, ansatz, hamiltonian, estimator):
    """Return estimate of energy from estimator

    Parameters:
        params (ndarray): Array of ansatz parameters
        ansatz (QuantumCircuit): Parameterized ansatz circuit
        hamiltonian (SparsePauliOp): Operator representation of Hamiltonian
        estimator (Estimator): Estimator primitive instance

    Returns:
        float: Energy estimate
    """
    cost = estimator.
        run(ansatz, hamiltonian, parameter_values=params).
        result().values[0]
    return cost

x0 = 2 * np.pi * np.random.rand(ansatz_ibm.num_parameters)

res = minimize(cost_func, x0, args=(ansatz, hamiltonian, estimator), 
    method="COBYLA",options={'disp': True})

# Assign solution parameters to ansatz
qc = ansatz.assign_parameters(res.x)
# Add measurements to our circuit
qc.measure_all()

# Sample ansatz at optimal parameters
samp_dist = sampler.run(qc).result().quasi_dists[0]
# Close the session since we are now done with it
session.close()
\end{lstlisting}

\section{Conclusion}

I got many interesting result in this project. First and foremost, this is the first time I test that the QAOA really works, for solving the NP complete problem such as hamiltonian cycle path problem. However, since the embedding require $(n-1)^2$ qubits, I can only simulate up to no more than $n=6$. 

In this problem, I choose $n=3,4$ and run the simulation on the simplest case: A triangle and a square. In both cases, I analyze the energy spectrum of the cost hamiltonian beforehand, and didn't start the simulation of QAOA until I'm convinced that the cost hamiltonian is correct. This step actually benefit me a lot in understanding the behavior of QAOA. One important thing is that once you know what exactly the minimum energy is, you can check the quality of QAOA parameter optimizer, by comparing the cost given by QAOA circuit with the minimum energy. Another interesting observation in the spectrum plotted in \autoref{fig:TriangleSpec} and \autoref{fig:FourEigen} is that in both cases \textbf{there is a large enough energy gap between the solution space the the non-solution space.} I highly doubt that, such energy gap is crutial for the success and time complexity of QAOA. There is no doubt that when the structure of the graph get more complicated, the spectrum can also get complicated, and thus it becomes harder for an optimizer to tell whether we are getting closer to ground state or not. On the other hand, if we could find a better general way of embedding Hamiltonian cycle, such that the solution space has a large gap between the non-solution space, than I would be much more confident that we can use QAOA to get the accurate solution state.

One thing one can never neglect is the role of quantum noise. Intuitively, when we add more noise into the circuit, our algorithm will only get worse result. The experiments of triangle case, in \autoref{fig:FakeManilaCompare}, is consistent with this intuition. However, in the square case, \textbf{the result of running QAOA is much better than on a noiseless simulator!} Does that mean, we don't need error correction at all for QAOA? Because noise seems to be a resource that benefit our optimization and annealing process, rather than a harmful factor! The idea of utilization quantum noise, is so crazy but attracting. Maybe I will explore this possiblity someday in the future.

\newpage

\printbibliography %Prints bibliography

\end{document}